\documentclass[12pt,preprint]{emulateapj}

\usepackage{graphicx}
\usepackage{amsmath}
\usepackage{hyperref}
\usepackage{color}

\hypersetup{
    colorlinks=true,
    linkcolor=red,
    citecolor=blue,
    urlcolor=magenta,
}

\newcommand{\gl}{G_{\rm light}}  
\newcommand{\be}{\begin{equation}}
\newcommand{\ee}{\end{equation}}
\newcommand{\bea}{\begin{eqnarray}}
\newcommand{\eea}{\end{eqnarray}}

\begin{document}

\title{Determining Model-Independent $H_0$ and Consistency Tests} 
\author{Kai Liao$^{1}$}
\author{Arman Shafieloo$^{2,3}$}\email{shafieloo@kasi.re.kr}
\author{Ryan E. Keeley$^{2}$}
\author{Eric V. Linder$^{2,4,5}$}
\affil{
$^1$ {School of Science, Wuhan University of Technology, Wuhan 430070, China}\\
$^2$ {Korea Astronomy and Space Science Institute, Daejeon 34055, Korea}\\
$^3$ {University of Science and Technology, Daejeon 34113, Korea}\\
$^4$ {Berkeley Center for Cosmological Physics and Berkeley Lab, University of California, Berkeley, CA 94720, USA}\\
$^5$ {Energetic Cosmos Laboratory Nazarbayev University, Nur-Sultan, Kazakhstan 010000}
}

\begin{abstract}
We determine the Hubble constant $H_0$ 
precisely (2.3\% uncertainty) in a manner independent of cosmological model through Gaussian process regression, using 
strong lensing and supernova data. 
Strong gravitational lensing of a variable source can provide a  time-delay distance $D_{\Delta t}$ and angular diameter distance to the lens $D_{\rm{d}}$. 
These absolute distances can anchor Type Ia supernovae, which give an excellent
constraint on the shape of the distance-redshift relation. 
Updating our previous results to use the H0LiCOW program's milestone dataset consisting of six lenses,
four of which have both $D_{\Delta t}$ and $D_{\rm{d}}$ measurements, 
we obtain $H_0=72.8_{-1.7}^{+1.6}\rm{\ km/s/Mpc}$ for a flat universe and $H_0=77.3_{-3.0}^{+2.2}\rm{\ km/s/Mpc}$ for a non-flat universe. 
We carry out several consistency checks on the data and find no statistically significant 
tensions, though a noticeable redshift dependence persists in a particular systematic manner 
that we investigate. 
Speculating on the possibility that this trend 
of derived Hubble constant with lens distance 
is physical, we show how this can arise through  
modified gravity light propagation, which would 
also impact the weak lensing $\sigma_8$ tension. 
\end{abstract}

\section{Introduction}
The flat cosmological constant plus cold dark matter ($\Lambda$CDM) model is currently taken as the concordance  cosmological scenario and explains a wide range of observations. However, there is significant tension in the $H_0$ value inferred within $\Lambda$CDM from the cosmic microwave background (CMB) observations $H_0=67.4 \pm 0.5\rm{\ km/s/Mpc}$~\citep{Planck2018} and 
that measured through the Cepheid distance ladder, $H_0=74.03\pm1.42 \rm{\ km/s/Mpc}$~\citep{Riess2019}. 
Other cosmological probes such as 
baryon acoustic oscillations together with primordial nucleosynthesis constraints agree with the CMB value of $H_0$ \citep{1707.06547,1811.02376,1906.11628,2002.04035}, while other distance ladder techniques can lie in between \citep{Freedman2020}. 

The discrepancy could arise due to unaccounted for systematic errors in observations or reveal new physics significantly different from $\Lambda$CDM. 

Independent cosmological probes could provide new perspectives. 
Strong gravitational lensing by galaxies offers an independent method of determining $H_0$ through time-delay lens systems. A typical lensing system consists of a distant active galactic nucleus (AGN) lensed by a foreground elliptical galaxy, forming multiple 
images along with the arcs of the host galaxy. The images will be magnified and the light will arrive at the Earth delayed by various times. Since AGNs are variable, we can measure the time delays between any two images from their light curves. 

From the time delay plus the lens potential measured by the high-resolution imaging and line-of-sight 
environment we can measure the ``time-delay distance" $D_{\Delta t}$. This is a combination of three angular
diameter distances, primarily depending on $H_0$ though also more weakly depending on the cosmological model, e.g.\ the matter density, dark energy properties, etc.  In addition, with kinematic information on the lens galaxy (for example the velocity dispersion) the angular diameter distance to the lens $D_{\rm{d}}$ can be also determined independent of the external convergence from line of sight  perturbers~\citep{Paraficz2009,Jee2015,Jee2016}. 

Either $D_{\Delta t}$ or  $D_{\rm{d}}$, or both jointly, 
provides a one-step method of determining $H_0$, which is substantially independent of and complementary to the CMB, large scale structure, and distance ladder methods. Thus they give a much needed crosscheck. However, like the other cosmological probes, one has to assign a cosmological model when computing the lensing distances. The results may therefore differ for  different models.

Rather than computing the distances within a model, one can instead {\it measure\/} the distance-redshift relation. To do so, one needs a probe that is both accurate and samples distance much more densely. Type Ia supernovae (SNe Ia) are superb mappers of the distance-redshift relation. However, they only provide relative distances because the SNe absolute magnitude (and Hubble constant) is unknown. The strengths of the two probes can combine together to remove each weakness: absolute distance measurements from time-delay lensing and relative distances from SNe Ia can anchor each other 
(for anchoring one type of distance with another, see for example  \citet{Aubourg2015,Cuesta2015,Collett2019,Pandey2019}). 
When combining the two, the results under different cosmological models seem to be stable and consistent~\citep{Taubenberger2019}.

One can make the cosmology-model independence (i.e. no form assumed for the expansion history $H(z)$) more explicit than simply checking under different models. In \citet{Liao2019} we applied Gaussian process (GP) regression to SNe Ia data to get a model-independent relative distance-redshift relation, i.e.\ the shape of distance-redshift function. Anchoring this together with $D_{\Delta t}$ from 4 H0LiCOW lenses resulted in  $H_0=72.2\pm2.1\rm{\ km/s/Mpc}$ in a flat universe and $H_0=73.0_{-3.0}^{+2.8}\rm{\ km/s/Mpc}$ considering the cosmic curvature density in the range $\Omega_{\rm{k}}=[-0.2,0.2]$. 

Currently, the H0LiCOW program has reached its first milestone~\citep{Wong2019}. The full dataset consists of 6 lenses, five of which were analyzed blindly, and four of which have both $D_{\Delta t}$ and $D_{\rm{d}}$ measurements.
In a flat $\Lambda$CDM cosmology, they found $H_0=73.3_{-1.8}^{+1.7}\rm{\ km/s/Mpc}$, consistent with the local Cepheid distance ladder measurement but in $3.1\sigma$ tension with $Planck$ and other CMB and large scale structure measurements. 
Note the H0LiCOW results vary with the assumed models~\citep{Wong2019}, although $H_0$ tends to increase 
with the usual generalized cosmologies. Furthermore, the 
previously identified trend \citep{Wong2019,Liao2019} in derived $H_0$ value, or time  delay distance excess, with redshift persists. Therefore, it is timely and useful to update our model-independent results. 

We briefly introduce the H0LiCOW program and the lensing dataset in Section~\ref{sec:lens}, and present our methodology and updated results in Section~\ref{sec:results}. In Section~\ref{sec:consis} we carry 
out some consistency checks between lensing data, and between 
lensing and supernova data. We explore the previously identified trend of $H_0$ with time-delay distance or lens redshift in Section~\ref{sec:grav}, and present a possible physical explanation based on modified gravity. 
We summarize and discuss the results and next steps in Section~\ref{sec:concl}.

\section{Lensing distances and H0LiCOW program} \label{sec:lens} 

Strong lensing (SL) by elliptical galaxies is a powerful tool to study both
astrophysics and cosmology~\citep{Treu2010}. Lenses with time-delay measurements were proposed to measure $H_0$~\citep{Refsdal1964,Treu2016}. Specifically, the time delay between any two images is determined by 
\begin{equation}
\Delta t=D_{\Delta t}\Delta \phi(\boldsymbol{\xi}_\mathrm{lens}),
\end{equation}
where the time-delay distance 
\begin{equation}
D_{\Delta t}=(1+z_\mathrm{d})\frac{D_\mathrm{d}D_\mathrm{s}}{D_\mathrm{ds}}
\end{equation}
is a combination of three angular diameter distances $D_{\rm{d}}, D_{\rm{s}}, D_{\rm{ds}}$, where the subscripts d and s 
denote the deflector (lens) and the source, respectively. 
We use units where $c=G=1$ throughout the article. 
The Fermat potential difference $\Delta \phi$ between the two images is a function of lens mass profile parameters $\boldsymbol{\xi}_{\rm{lens}}$,  determined by high-resolution imaging of the host arcs. Note that all other mass along the line of sight could also contribute to the lens potential, causing additional (de)focusing of the light rays and affecting the observed time delays. Considering the
effects of the perturber masses are small, they can be approximated by an external convergence $\kappa_{\rm{ext}}$. Then the inferred $D_{\Delta t}$ will be scaled by 1-$\kappa_{\rm{ext}}$. Independent observations such as galaxy counts could break the degenaracy~\citep{Rusu2017}. We take the time delay distances as given by H0LiCOW, and refer readers to the systematics treatment in \cite{Millon2019}.

Additional information on the lens galaxy such as the light profile $\boldsymbol{\xi}_{\rm{light}}$, the projected stellar velocity dispersion $\sigma^{\rm{P}}$, and the anisotropy distribution of the stellar orbits, parametrized by  $\beta_{\rm{ani}}$, can yield the angular diameter distance to the lens  \citep{Birrer2016,Birrer2019}: 
\begin{equation}
D_\mathrm{d}=\frac{1}{1+z_\mathrm{d}}D_{\Delta t}\frac{J(\boldsymbol{\xi}_\mathrm{lens},\boldsymbol{\xi}_\mathrm{light},\beta_\mathrm{ani})}{(\sigma^\mathrm{P})^2}\,,
\end{equation}
which correlates with $D_{\Delta t}$. The function $J$ captures all the model components computed from
angles measured on the sky (from imaging) and the stellar orbital anisotropy distribution (from spectroscopy). 
We refer to Section~4.6 of \citet{Birrer2019} for the detailed modeling related to $J$.

The state-of-the-art lensing collaboration H0LiCOW aims at measuring $H_0$ with $1\%$ precision using time-delay lenses~\citep{Suyu2017}. They take advantage of substantial data consisting of time-delay measurements from
the COSMOGRAIL program\footnote{http://www.cosmograil.org} and radio-wavelength monitoring, deep HST
and ground-based adaptive optics imaging, spectroscopy of the lens galaxy, and deep wide-field spectroscopy and imaging. 

In the recent milestone paper~\citep{Wong2019}, 
they gave the latest constraints on $H_0$ under different cosmological models
with a combined sample of six lenses that span a range of lens and source redshifts,
as well as various image configurations (double, cross, fold, and cusp).  
All lenses except the earliest, B1608+656, were analyzed blindly with respect to the cosmological parameters.
Four lenses (RXJ1131-1231, PG 1115+080, B1608+656, SDSS 1206+4332) have both $D_{\Delta t}$
and $D_{\rm{d}}$ measurements. 
Note that $D_{\Delta t}$ and $D_{\rm{d}}$ measurements for a system are correlated except for B1608+656 whose distance measurements
are independent.
The distance posterior distributions are released in the form of MCMC chains or skewed log-normal function fits
on the H0LiCOW website\footnote{http://www.h0licow.org}. 
For the case of MCMC chains, we will get the likelihood functions by smoothing the discrete points.
We summarize the redshifts and measured distances in Table~\ref{redshifts} ordered
by the lens redshift. For more detailed information on these lenses, see \citet{Wong2019} and the references therein.

\begin{table*}\centering
\begin{tabular}{cllllll}
\hline\hline
Order & Name & $z_{\rm{d}}$ & $z_{\rm{s}}$ & $D_{\Delta t}$ (Mpc) & $D_{\rm{d}}$ (Mpc) & references\\
\hline
\rule{0pt}{1.1\normalbaselineskip}1 & RXJ1131-1231 & 0.295 & 0.654 & $2096_{-83}^{+98}$ & $804_{-112}^{+141}$ & (1)(2)\\
\rule{0pt}{1.1\normalbaselineskip}2 & PG 1115+080  & 0.311 & 1.722 & $1470_{-127}^{+137}$ & $697_{-144}^{+186}$ & (1)\\
\rule{0pt}{1.1\normalbaselineskip}3 & HE 0435-1223 & 0.4546 & 1.693 & $2707_{-168}^{+183}$ & -  &  (1)(3)\\
\rule{0pt}{1.1\normalbaselineskip}4 & B1608+656 & 0.6304 & 1.394  &  $5156_{-236}^{+296}$ & $1228_{-151}^{+177}$  & (4)(5)\\
\rule{0pt}{1.1\normalbaselineskip}5 & WFI2033-4723 & 0.6575 & 1.662 & $4784_{-248}^{+399}$ & - & (6)\\
\rule{0pt}{1.1\normalbaselineskip}6 & SDSS 1206+4332 & 0.745 & 1.789 & $5769_{-471}^{+589}$ & $1805_{-398}^{+555}$ & (7)\\ 
\hline\hline
\end{tabular}
\caption{Redshifts and distances of the six H0LiCOW lenses ordered by lens redshift.
The references are (1) \citet{Chen2019}, 
(2) \citet{Suyu2014}, (3) \citet{Wong2017}, 
(4) \citet{Suyu2010}, (5) \citet{Jee2019}, 
(6) \citet{Rusu2019}, (7) \citet{Birrer2019}. 
}\label{redshifts}
\end{table*}

\section{Methodology and results} \label{sec:results} 

To combine the SNe data with lensing data, we generate samples of unanchored luminosity distance $H_0D^{\rm{L}}$ from the posterior of the Pantheon compilation from ~\citet{pantheon}, calculated with a GP (See, e.g., \cite{2001.10887} for a test of cosmology model independence.) This paper follows the analysis in \citet{Liao2019}, which is based on the \texttt{gphist}~\citep{GPHist} code first presented in \cite{Keeley0}.  

Regression using a GP works by generating a set of functions from an infinite dimensional function space characterized by a covariance function.  This covariance function is parametrized by a squared-exponential kernel
\begin{equation}
    \langle \gamma (s_1) \gamma (s_2) \rangle = \sigma_f^2 \exp\left[-(s_1-s_2)^2/(2\ell^2) \right]\ ,
\end{equation} 
where $s_i=\log(1+z_i)/\log(1+z_{\rm max})$ and $z_{\rm max}=2.26$ is the maximum redshift of the supernova sample. 
This has two hyper-parameters that are marginalized over, $\sigma_f$ and $\ell$, with $\sigma_f$ determining the amplitude of the random fluctuations and $\ell$ determining the coherence length of the fluctuation, equivalently $1/\ell$ is proportional to the number of fluctuations in the range.  The priors on these hyper-parameters are scale-invariant and we directly integrate over this space since the dimensionality is small.

The GP function $\gamma(z) = \ln (H^{\rm fid}(z)/H(z))$ involves the expansion history $H(z)$ (which can be integrated to give distances), and $H^{\rm fid}(z)$ is the best-fit $\Lambda$CDM expansion history from the Pantheon dataset and works as the mean function of the GP regression. The GP prior functions are then trained on the Pantheon likelihood, which constrains only the shape of the expansion history, not the absolute scale, so unanchored luminosity distances $H_0D^{\rm{L}}(z)$ are the quantities most directly constrained by the Pantheon SNe dataset.

\begin{table*}\centering
 \begin{tabular}{lccccc}
  \hline\hline 
Case &   $6\,D_{\Delta t}+ 4\,D_{\rm{d}}$  &  $6\,D_{\Delta t}$  & $4\,D_{\rm{d}}$ & $4\,D_{\Delta t}$ (Liao et al.\ 2019) & $6\,D_{\Delta t}+4\,D_{\rm{d}}$ (non-flat)\\

\rule{0pt}{1.1\normalbaselineskip}$H_0$ ($\rm{km/s/Mpc}$)   & $72.8_{-1.7}^{+1.6}$ & $72.8_{-1.8}^{+1.7}$ & $81.0_{-6.9}^{+7.1}$ & $72.2\pm2.1$ & $77.3_{-3.0}^{+2.2}$\\
  \hline\hline
 \end{tabular}
 \caption{Median values plus 16th and 84th percentiles for data combinations.
}\label{stats}
\end{table*}

\begin{table*}\centering
 \begin{tabular}{lcccccc}
  \hline\hline
Lens &   RXJ1131-1231 &  PG 1115+80 & HE0435-1223 & B1608+656 & WFI2033-4723 & SDSS 1206+4332\\

\rule{0pt}{1.1\normalbaselineskip}$H_0$ ($\rm{km/s/Mpc}$)   & $77.5_{-3.5}^{+3.4}$ & $80.5_{-7.1}^{+8.1}$ & $71.0_{-4.4}^{+4.8}$ & $70.5_{-3.2}^{+2.9}$ & $71.8_{-4.7}^{+3.8}$ &  $67.9_{-4.8}^{+5.3}$\\
  \hline\hline
 \end{tabular}
 \caption{Median values plus 16th and 84th percentiles for each lens system. The combined $D_{\Delta t}$ and $D_d$ (if it has $D_d$) data are used. The universe is set to be flat.
}\label{stats_each}
\end{table*}

To summarize the method for determining $H_0$:
\begin{enumerate}
\item
Draw 1000 unanchored luminosity distance curves $H_0D^{\rm{L}}$ from the GP fit to the SNe data, and convert
to unanchored angular diameter distances $H_0D^{\rm{A}}$;
\item Evaluate the values of each of the 1000 $H_0D^{\rm{A}}$ curves at the lens and source redshifts of the six (SL) systems to calculate 1000 values of $H_0D_{\Delta t}$ for each system
using 
\be 
H_0D_{\Delta t}=(1+z_\mathrm{d})(H_0D_\mathrm{d})(H_0D_\mathrm{s})/(H_0D_\mathrm{ds})\,; 
\ee 
\item Compute the likelihood, for each of the 1000 realizations, from the H0LiCOW's $D_{\Delta t}$ combined with $D_{\rm{d}}$ data (if the $D_{\rm{d}}$ measurements are available) 
for each lens system for many values of $H_0$;
\item Multiply the six likelihoods to form the full likelihood for each realization, for each value of
$H_0$;
\item Marginalize over the realizations to form the posterior distribution of $H_0$.
\end{enumerate}

Note that to obtain the angular diameter distance $D_{\rm{ds}}$ between the lens and the source  from $H_0D^{\rm{A}}$ in step 2, we use the standard distance relation \citep{Weinberg1972} 
\begin{equation}
\begin{split}
D_\mathrm{ds}&=D_s\sqrt{1+\Omega_\mathrm{k}(1+z_\mathrm{d})^2(H_0D_\mathrm{d})^2}  \\
&\qquad-\frac{1+z_\mathrm{d}}{1+z_\mathrm{s}}D_\mathrm{d}\sqrt{1+\Omega_\mathrm{k}(1+z_\mathrm{s})^2(H_0D_\mathrm{s})^2} \ ,
\end{split}
\end{equation}
where $\Omega_k$ is the dimensionless curvature density.
For a spatially flat universe,
one simply has $D_{\rm{ds}}=D_{\rm{s}}-[(1+z_{\rm{d}})/(1+z_{\rm{s}})]\,D_{\rm{d}}$. 

In step 3, we also evaluate a case with the SL $D_{\rm{d}}$ data alone. In such a case, the anchoring is direct between absolute and relative distances and one does not need to consider the assumption of cosmic curvature.

\begin{figure}
\includegraphics[width=\columnwidth,angle=0]{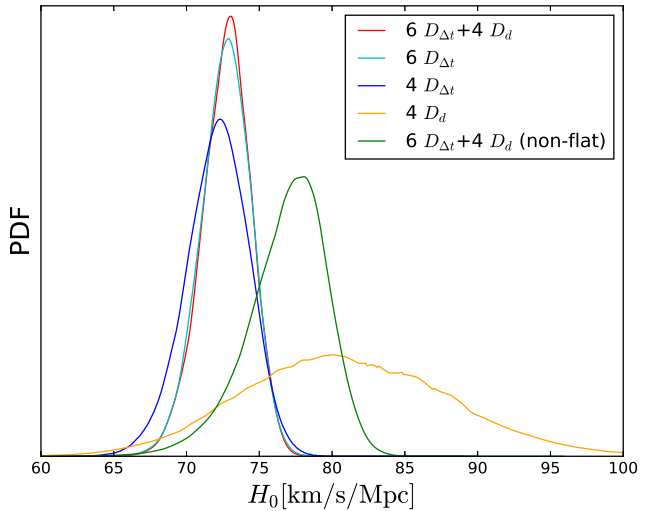}
   \caption{The joint posteriors on $H_0$ are shown for the 
   full data set ``6 $D_{\Delta t}$ + 4 $D_{\rm{d}}$'', the 
   time-delay distances ``6 $D_{\Delta t}$'' and angular distances 
   ``4 $D_{\rm{d}}$'' separately, and the change from the previous 
   ``4 $D_{\Delta t}$'' of \cite{Liao2019}. Here we 
   assume flatness to compute $D_{\Delta t}$ from the SNe GP by default.
   For completeness, we also consider the non-flat case for the full data set. 
  }\label{results}
\end{figure}

\begin{figure}
\includegraphics[width=\columnwidth,angle=0]{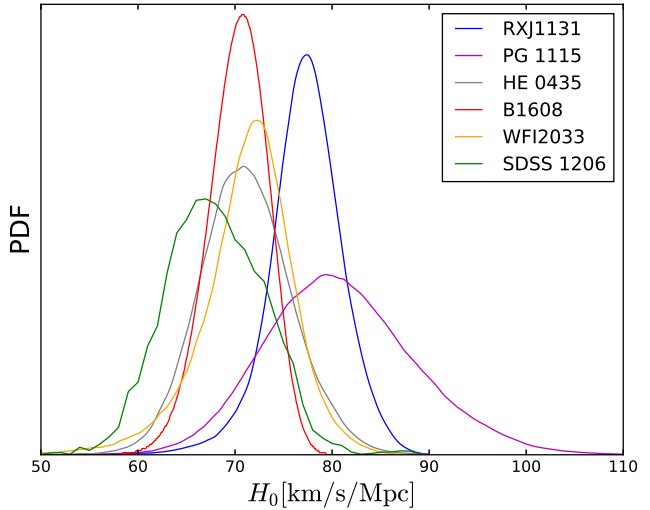}
   \caption{The individual posteriors on $H_0$ from each 
   lensing system are shown, for flat cosmology. 
  }\label{each}
\end{figure}

Figure~\ref{results} shows the joint posteriors and
the numerical results are summarized in Table~\ref{stats}.
First of all, we consider the full data set consisting 
of 6 lenses, 4 of which have both $D_{\Delta t}$ and $D_d$ measurements.
The constraint is $H_0=72.8_{-1.7}^{+1.6}\rm{\ km/s/Mpc}$ assuming the universe is flat.
It is more stringent than our previous result based on 4 lenses with only $D_{\Delta t}$ data, 
$H_0=72.2\pm2.1\rm{\ km/s/Mpc}$. 
This model-independent result has comparable value and error bars as that under the flat $\Lambda$CDM model by H0LiCOW, $H_0=73.3_{-1.8}^{+1.7}\rm{\ km/s/Mpc}$, 
supporting the tension with the BAO and CMB results. 
(Note that the looser constraints expected from removing the $\Lambda$CDM assumption 
are offset by including supernova distances.) 

In addition, we test the contribution of the $D_{\rm{d}}$ data by taking $D_{\Delta t}$ and $D_{\rm{d}}$ separately. 
The constraint from 4 $D_{\rm{d}}$ data alone is relatively weaker,  $H_0=81.0_{-6.9}^{+7.1}\rm{\ km/s/Mpc}$, 
and the 6 $D_{\Delta t}$ data show almost the same constraint power, $H_0=72.8_{-1.8}^{+1.7}\rm{\ km/s/Mpc}$, as the full set.
Nevertheless, the result from $D_{\rm{d}}$ data is free from 
assumptions concerning spatial curvature (although note that other lens parameters, such as stellar anisotropy, enter.) 

We also consider the non-flat case for the full dataset for completeness. We set the uniform prior of the curvature parameter to be $\Omega_{\rm{k}}=[-0.5,0.5]$ as in Wong et al. (2019). The constraints are $H_0=77.3_{-3.0}^{+2.2}\rm{\ km/s/Mpc}$ and $\Omega_{\rm{k}}=0.33_{-0.19}^{+0.12}$. Like the $\Lambda$CDM-assumed case where $H_0=74.4_{-2.3}^{+2.1}\rm{\ km/s/Mpc}$,
$\Omega_{\rm{k}}=0.26_{-0.25}^{+0.17}$~\citep{Wong2019},  our non-flat result shows a larger $H_0$ (possibly the $D_d$ data, 
which give a higher $H_0$, have more influence in this case) and slightly favors an open universe. However, the change of $H_0$ relative to the flat case in this work is more distinct.
It is worth mentioning that the constraint on $\Omega_{\rm{k}}$ in our method is model-independent as well. 

Furthermore, to understand the relative contributions, we  constrain $H_0$ with each lens in the model-independent manner.
We use both $D_{\Delta t}$ and $D_{\rm{d}}$ data (if it has). 
Figure~\ref{each} shows the individual posteriors. The numerical results are shown in Table~\ref{stats_each}, in order of increasing lens redshift. As one can see, a trend may exist: the $H_0$ values roughly decrease with the lens redshift
or the distance. Our results further confirm the trend noticed in \citet{Wong2019,Liao2019} and shown in Fig.~5 in \citet{Millon2019} based on $\Lambda$CDM.
We explore this further in Section~\ref{sec:grav}.

\section{Consistency tests} \label{sec:consis}

In this section, we check to what extent the SL data are internally consistent and to what extent they are consistent with the SNe data. This is important to confirm that 
the results from the combination of lens systems, and from 
combination of SL and SNe, are robust. 
As the available SL distances become more numerous and more precise, one can better assess the consistency of these distances with the SNe distances.  We begin by comparing the distance posteriors of the SL data, with the values of those quantities predicted by the SNe data. 

\begin{figure*}
    \centering
    \includegraphics[width=\textwidth]{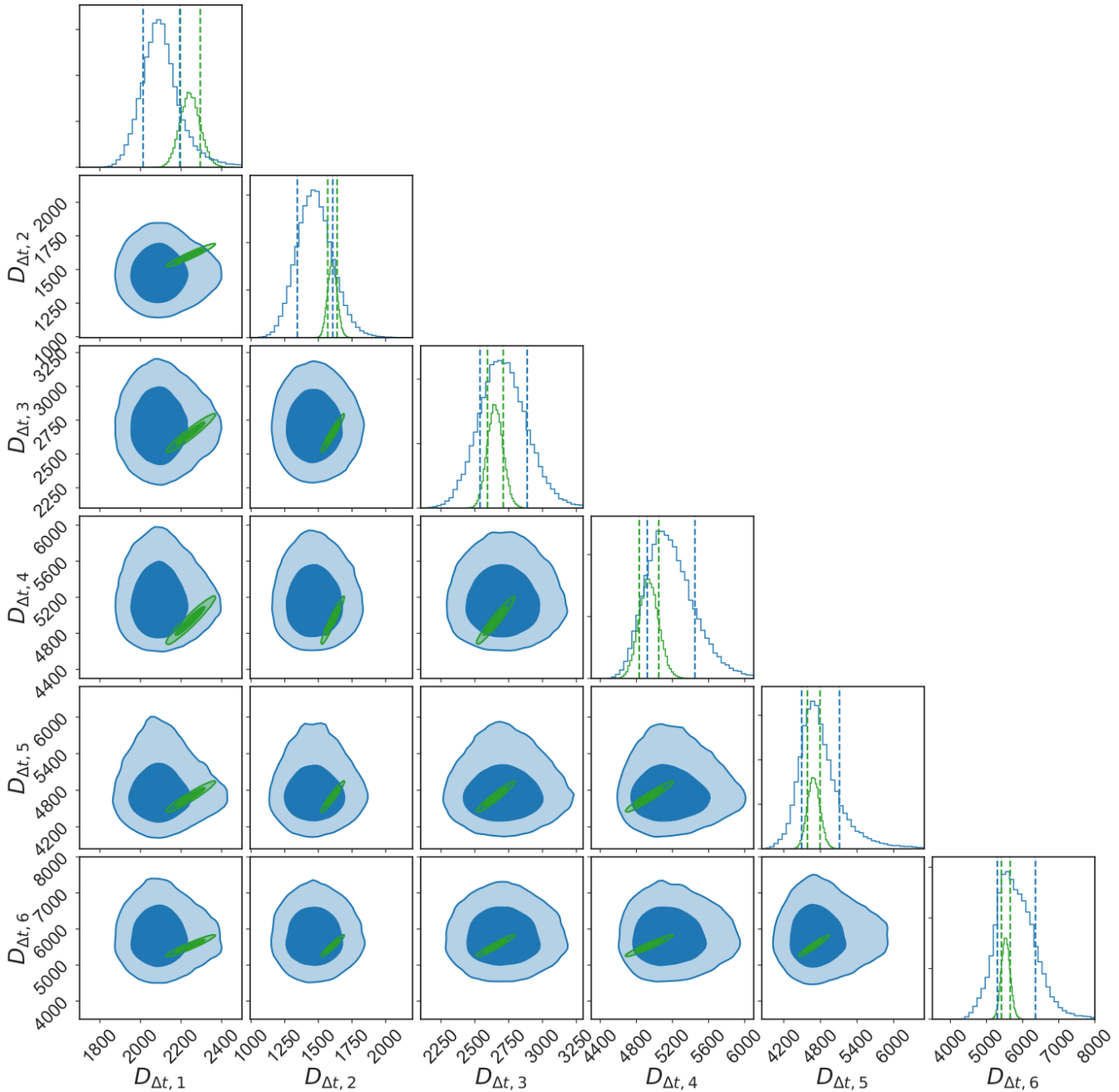}
    \caption{
2D contours of the posteriors, and 1D marginalized probability distribution functions, of the SL time-delay distances (blue) and the posterior sampled distances calculated from the GP reconstruction from SNe (green), at 68\% and 95\% confidence level. The units are Mpc. The major axis of the green SNe contours corresponds to variation in the value of $H_0$. Systems are ordered by redshift (see Table~\ref{redshifts}). 
    }
    \label{fig:jackknife}
\end{figure*}

In Fig.~\ref{fig:jackknife}, we plot the measured posteriors of the individual systems' time-delay distances at the 68\% and 95\% confidence level, as well as the SNe posterior predictions of those systems' time-delay distances using a GP regression on the Pantheon data. (The uncertainties on the angular distances are 
currently too large to add useful information to this comparison.) Since the SNe cannot constrain an absolute scale, the SNe posterior predictions for the SL distances are anchored in this figure by our combined $H_0$ measurement $H_0=72.8^{+1.6}_{-1.7}\rm{\ km/s/Mpc}$, giving the extended  
green contours, reflecting both the uncertainty in $H_0$ and the GP-based unanchored distances ($H_0 D_A$).  On the whole, the measurements of the SL distances are consistent with the SNe posterior predictions, in  
both the 1D and 2D joint posteriors, although since we have 
ordered the systems by lens redshift one can notice  
a tendency for the SL posterior to drift rightward (to higher 
distance and hence lower $H_0$) across the SNe posterior with  
increasing redshift. This will be explored further in Sec.~\ref{sec:grav}.

\begin{figure*}
    \centering
    \includegraphics[width=\columnwidth]{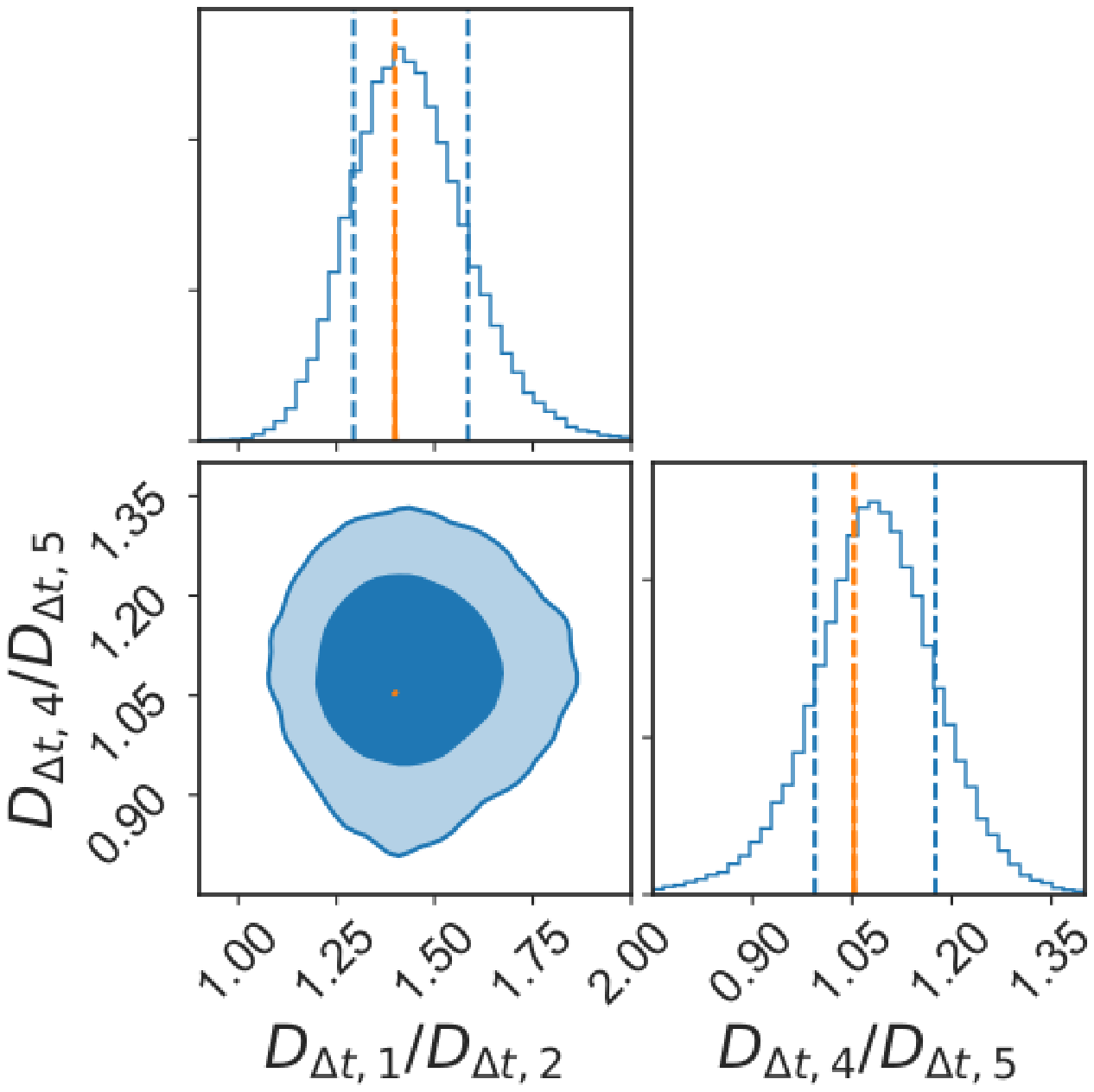}
    \includegraphics[width=\columnwidth]{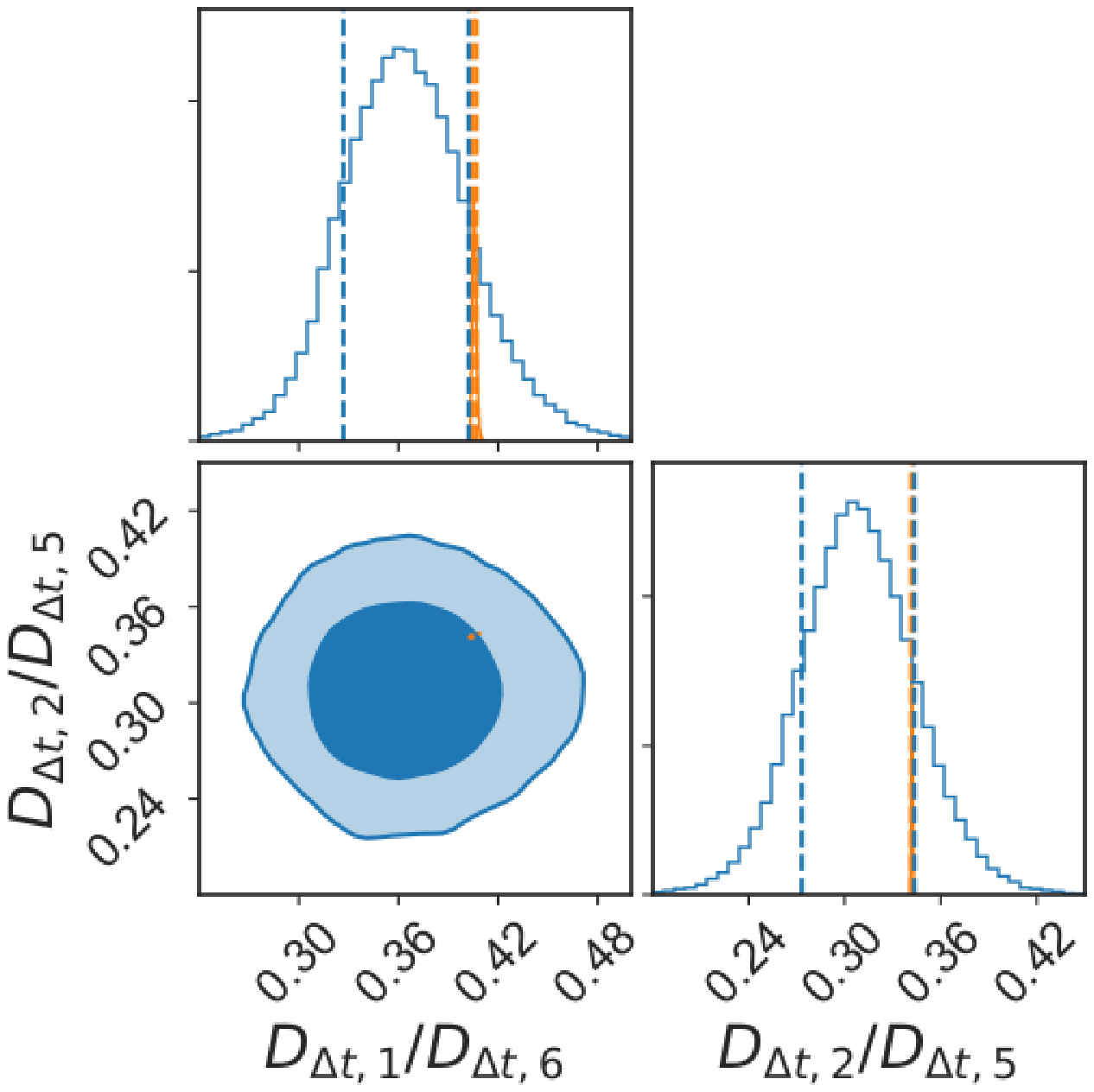}
    \caption{Corner plots of time-delay distance ratios, hence independent of Hubble constant, with SL results (blue) and GP reconstruction from SNe (orange). The left panel is for systems at almost the same redshift (whether lower or higher than $z_{\rm{d}}=0.4$), with lenses and SNe in excellent agreement,  while the right panel is for systems at very different redshifts (one lower, one higher than $z_{\rm{d}}=0.4$), where the comparison pulls to the edge of the 68\% confidence  contour. Note the tight constraints on distance ratio (i.e.\ relative distance) from SN, giving a very small orange contour.
    }
    \label{fig:ratio1v2}
\end{figure*}

By forming ratios of time-delay distances, we can cancel out 
the dependence on $H_0$, forming relative distances that the 
SNe are particularly suited to. In Fig.~\ref{fig:ratio1v2}, we plot the measured SL posteriors of the ratios of time-delay distances for certain combinations of the systems, as well as the posterior predictions for those ratios from the SNe data.  Indeed, the SNe posterior predictions are nearly pointlike. Rather than 
showing hundreds of 2D joint combinations of the 15 possible 
time-delay ratios, we select two cases: the left panel shows 
the ratios for the two pairs of systems nearly at the same lens 
redshift (see  Table~\ref{redshifts}), and the right panel for 
two pairs of systems with the most extreme differences in lens 
redshift. The pairs nearly at the same redshift (one pair  
with both $z_{\rm{d}}<0.4$ and one pair with both $z_{\rm{d}}>0.4$) are highly 
consistent with the very precise SNe predictions. Those pairs 
at very different redshifts (thus one system has $z_{\rm{d}}<0.4$ and  
one system has $z_{\rm{d}}>0.4$), are shifted to the boundary of the 
68\% confidence region, in both 1D and 2D posteriors. This is 
not statistically significant, but does seem to continue a trend.

\begin{figure*}
    \centering
    \includegraphics[width=\columnwidth]{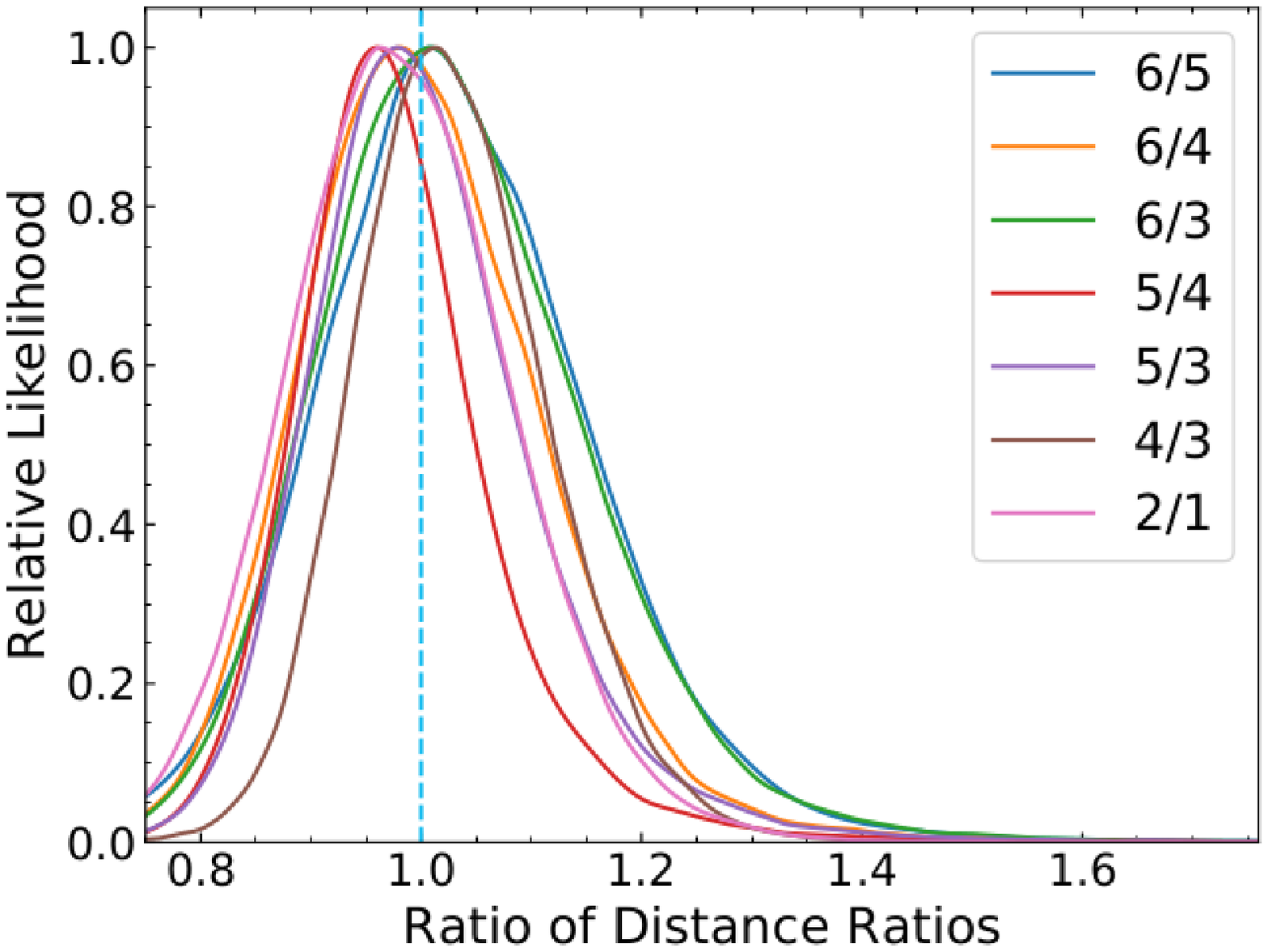}
    \includegraphics[width=\columnwidth]{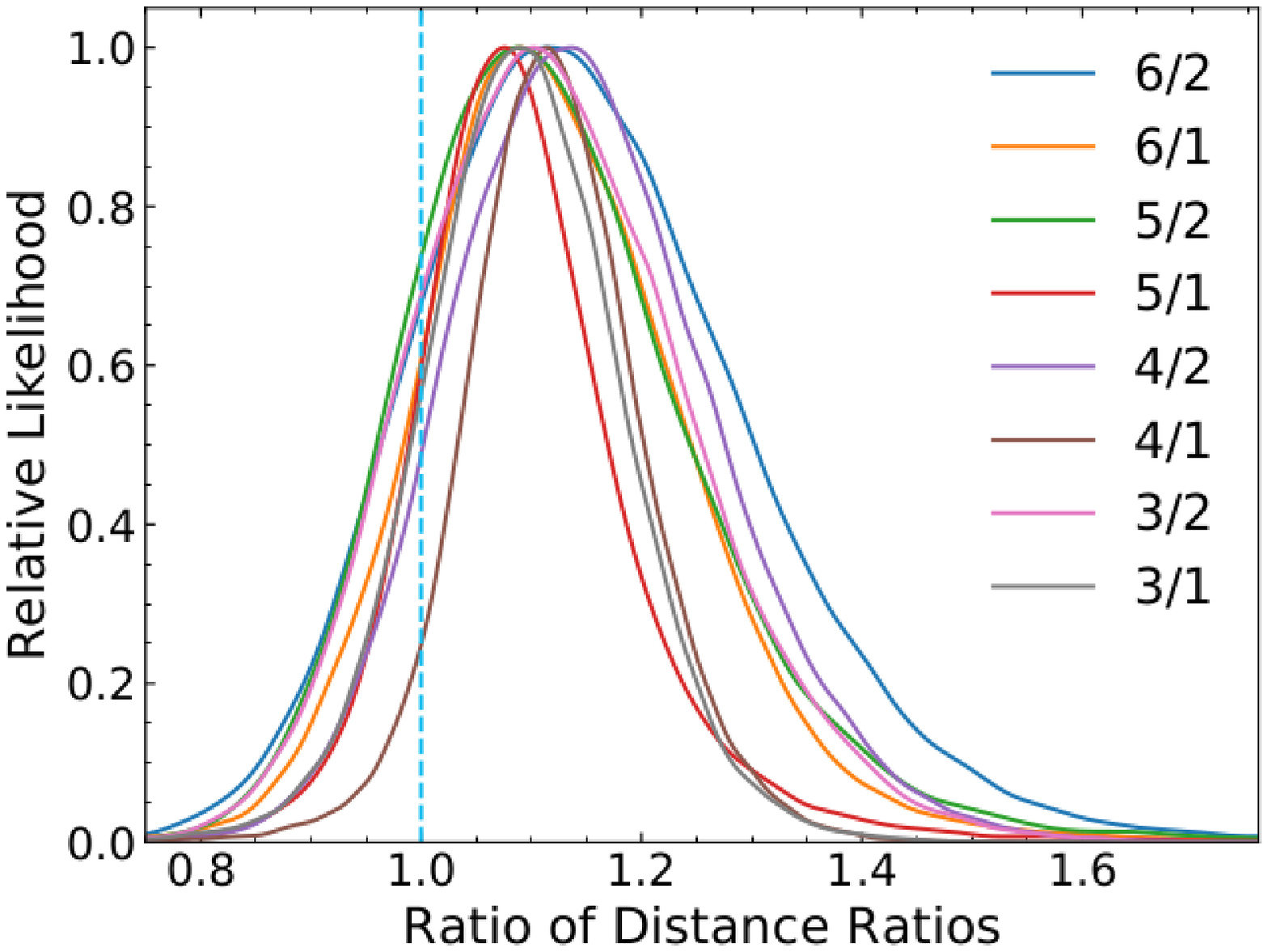}
    \caption{Relative likelihoods are plotted for the posteriors of ratios (SL to SNe) of distance ratios (hence independent of $H_0$). The label ``$i$'' means  $D_{\Delta t,i}/(D_{\Delta t,i})_{\rm SN}$. The vertical dotted line at unity shows the expectation if SL are consistent with SNe cosmology (independent of $H_0$). 
[Left panel] The ratios of ratios with both lenses above or both below $z_{\rm{d}}=0.4$ have posteriors peaking very close to unity, showing excellent consistency. [Right panel]  As the left panel, but now with one lens above and one lens below $z_{\rm{d}}=0.4$. While the posteriors are still consistent with unity, every one of the 8 possible combinations peaks above 1. 
    }
    \label{fig:ratioratio}
\end{figure*}

To investigate this further, 
in Fig.~\ref{fig:ratioratio}, we plot the measured SL 1D posteriors of every combination of the ratios of the time-delay distances of sources with lens redshifts both at $z_{\rm{d}}>0.4$ or both at $z_{\rm{d}}<0.4$ (left panel), or one at $z_{\rm{d}}>0.4$ and one at $z_{\rm{d}}<0.4$ (right panel). These 1D posterior ratios are given relative to the predictions of those same ratios from the SNe data, i.e.\ $(D_{\Delta t,i}/D_{\Delta t,j})/(D_{\Delta t,i}/D_{\Delta t,j})_{\rm SN}$, 
so consistency gives the value of 1. 
This can also be viewed equivalently as $(D_{\Delta t,i}/(D_{\Delta t,i})_{\rm SN})/(D_{\Delta t,j}/(D_{\Delta t,j})_{\rm SN}))$, testing that SL cosmology is consistent with SNe cosmology, irrespective of $H_0$. 

Interestingly, when both lens systems in 
the ratio have $z_{\rm{d}}>0.4$, or both have 
$z_{\rm{d}}<0.4$, then the posteriors of the 7 
possible combinations all peak very close 
to unity, showing consistency (see left 
panel). However, when the lenses in the 
ratio lie on opposite sides of $z_{\rm{d}}=0.4$, 
then although the posteriors are still consistent with the value 1, the peaks 
tend to be $\sim10\%$ higher. 
Any one posterior could 
fluctuate above the consistency value 
of 1, but we note that all 8 possible combinations all lie high. This is not to 
say that one should multiply 8 $1\sigma$ deviations (\citet{Wong2019} find this trend to be somewhat under $2\sigma$); we merely consider it odd 
enough to make it worthwhile looking for physics (remember, this 
is independent of the value of $H_0$) that could account for this. This is what we explore in the next section. Alternately, 
it could be due to some unidentified observational 
systematic, and one should see if this 
putative trend persists with new lens systems and new data.

\section{A Question of Gravity} \label{sec:grav} 

In going from the four lens systems of \cite{Birrer2019} to the 
seven lens systems of \cite{Wong2019} and \cite{Shajib2019} 
(the posterior chains for the last, DES J0408-5354, have not been publicly released yet but we include in this section their 
quoted $H_0$ value from it), the 
apparent trend of the time-delay distance with respect to the 
distance relation predicted by the supernova luminosity distance 
\citep{Liao2019}, or derived Hubble constant \citep{Wong2019}, 
as a function of time-delay distance or lens redshift has remained. 
In particular, see Fig.~5 of \citet{Millon2019}. 
The statistical sample is still small, so perhaps this can be just an 
odd fluctuation despite the extra systems not reducing the trend. 

In this section we briefly explore the conjecture that the trend is 
physical, and that it could be 
related not to the background expansion history (distances, so 
SN and BAO are unaffected) but rather the 
behavior of gravity on light deflection (lensing) evolving with redshift. 
This is commonly called $\gl(z)$ and arises in many modified 
gravity theories. 

The light deflection depends on the sum of the time-time and 
space-space metric potentials, $\Phi+\Psi$, and is related to 
the density contrast $\delta\rho/\rho$ by  
\be 
\nabla^2(\Phi+\Psi)=8\pi G_N\,\gl\,(\delta\rho/\rho)\  . \label{eq:poi} 
\ee 
Could the trend be reflecting $\gl(z)$? Note that $\gl$  
does not affect supernova distances, so those would reflect the 
actual background expansion history. 

Here we simply give a rough analysis, ignoring some subtleties 
we mention later. The measured lensing time delays depend on 
\be  
\Delta t= D_{\Delta t} \Delta\phi\ , 
\ee 
where $\Delta\phi$ is the difference in Fermat potentials. The 
Fermat potential difference 
\be 
\Delta\phi=[(\theta_2-\beta)^2-(\theta_1-\beta)^2]/2-[\psi(\theta_2)-\psi(\theta_1)]\ , 
\ee 
where $\theta_i-\beta$ is the angular difference between the image 
location and unlensed source location, and $\psi$ is a projected 
potential. The angular deflection with modified gravity light  
propagation becomes 
\be 
\vec\alpha=\vec\nabla\psi(\vec\theta)\quad\rightarrow\quad \vec\nabla[\gl\psi(\vec\theta)]\ , 
\ee 
and the projected potential is related to the convergence from 
mass along the line of sight by 
\be 
\nabla^2\psi(\vec\theta)=2\kappa(\vec\theta) \quad\rightarrow\quad 2\gl\kappa(\vec\theta)\  . 
\ee 

Thus the gravitational effects on light propagation give 
\be 
\Delta t= D_{\Delta t} \Delta\phi \quad\rightarrow\quad \Delta t= D_{\Delta t} \gl\,\Delta\phi\ . 
\ee 
Since the Hubble constant estimated from a lens system comes from 
$1/D_{\Delta t}$ then for given measured lens system 
characteristics we have 
\be 
H_0({\rm measured\ at\ }a)=H_{0,{\rm true}}\,\gl(a)\ . 
\ee 
That is, if $\gl$ is increasing with $a$ then the derived value of 
$H_0$ will increase for lower redshift lens systems. 

\begin{figure}[!tbhp] 
\centering
\includegraphics[width=\columnwidth]{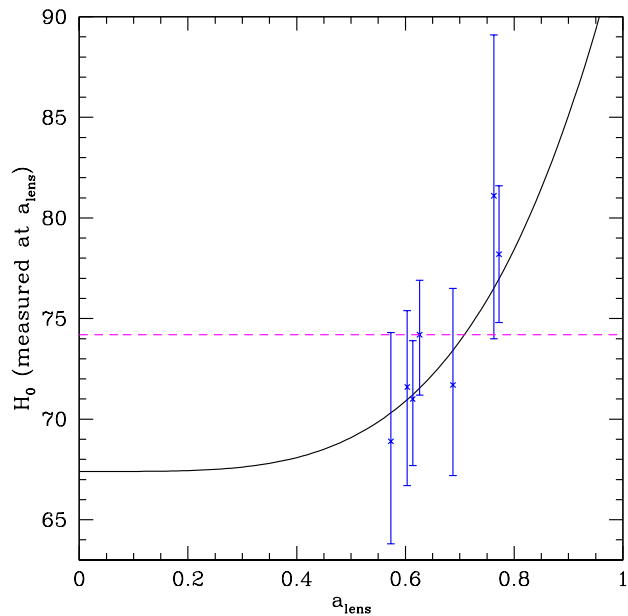} 
\caption{
Data points show the measured trend in derived Hubble constant 
$H_0$ measured from lenses at scale factor $a_{\rm lens}$, 
while the curve is the modified 
gravity toy model, Eq.~(\ref{eq:toy}), for measurements 
involving light deflection (which for $a_{\rm lens}\ll1$ goes 
to the true expansion rate $H_0$). The horizontal dashed magenta 
line shows the value of 
$H_0$ quoted by \citet{Millon2019} for the data. 
} 
\label{fig:glight}
\end{figure}

We can assume this is the cause of the observed trend in derived 
$H_0$ from the lens systems and derive what function $\gl(a)$ is needed. At high redshift 
we expect gravity to restore to general relativity (e.g.\ to 
preserve the successes of the cosmic microwave background and 
primordial nucleosynthesis) so we take values of $H_0$ derived 
at high redshift to be the true values. Figure~\ref{fig:glight} 
shows the $H_0$ values given in \citet{Millon2019}, within 
$\Lambda$CDM, for the 
seven lens systems 
and an illustrative power law in scale factor, 
\be 
\gl(a)=1+0.4a^4\ . \label{eq:toy} 
\ee 
(Note that for many modified gravity models $\gl$ actually 
levels out to a constant de Sitter value not far into the future.)

The rise in measured $H_0$ values comes in quite steeply with 
scale factor, around $a\gtrsim0.7$ (or $z\lesssim0.4$). A change 
so rapid to distances themselves is quite difficult to achieve, 
since distances are integrals over the expansion rate (and double 
integrals over dark energy equation of state). And certainly no 
such dramatic change in seen in supernova distances. However, by 
changing the gravitational strength affecting light deflection, 
$\gl(a)$, this is less difficult. Indeed, such a rapid change has 
been shown to occur for some actual modified gravity theories -- see 
for example Fig.~5 (right panel, thin curves) of \citet{Linder2017}. 
Note the numerical solution there shows that indeed the effect on 
$\gl(a)$ can first become significant at low redshifts. While 
that particular theory (uncoupled Galileon) is ruled out, it does 
give a proof of principle that modified gravity can act in such a 
manner. 

One might also speculate that the effect on light deflection could 
show up in weak lensing measurements. For the convergence or shear 
power spectrum, the relativistic Poisson equation (\ref{eq:poi}) 
shows that $\gl\kappa\sim(\Phi+\Psi)$ and so the measured shear 
power will be proportional to $\gl^{-2}$. There is an extra element 
in that the growth of structure also depends on $G_{\rm matter}$, 
but we focus here on $\gl$. The value of the mass fluctuation 
amplitude $\sigma_8$, or $S_8=\sigma_8(\Omega_m/0.3)^{1/2}$, 
derived from the shear power spectrum will thus be proportional 
to $1/\gl(a)$. This was discussed in \citet{Daniel2010} -- 
``For higher [$\gl$], lower values of $\sigma_8$ will produce 
comparable lensing potentials. Larger [$\gl$] does not cause 
$\sigma_8$ to decrease per se, rather it brings lower values of $\sigma_8$ 
into better agreement with the data''. Since we take $\gl$ to be 
strengthening at lower 
redshift for the SL case, this means that the value 
of $\sigma_8$ or $S_8$ derived from low  redshift surveys should 
be less than from high redshift surveys (or Planck CMB). This 
trend does seem consistent with weak lensing survey data (but 
again, evolution of $G_{\rm matter}$ can overturn this).  

Regarding the subtleties we mentioned at the beginning of this 
section, note that the light deflection occurs all along the 
path and not just at the lens, but as with general relativistic 
light deflection one can treat the deflection as occurring at the 
lens, in a single screen approximation. Right at the 
lens we might expect the modified gravity to be screened, but 
further out from the lens the screening vanishes and the dominant 
part of the path integral is roughly at the lens redshift. Thus we take 
$\langle \gl(a)\Delta\phi\rangle\approx \gl(a_{\rm lens})\Delta\phi$. 

Finally, it is not clear how $\gl(a)$ could help with the 
Cepheid measurement of higher $H_0$, though the value from the 
tip of the red giant branch technique is more consistent. 
This whole section is 
simply speculation, but if the trend in $H_0$ measurements 
with distance persists, we might consider it a question of 
gravity.

\section{Conclusion and Discussions} \label{sec:concl} 

Based on the method we previously proposed, we give a cosmology model-independent 
determination of $H_0$ with the updated H0LiCOW dataset consisting of 
six lenses. The absolute lensing distances ($D_{\rm{{\Delta t}}}$ and $D_{\rm{d}}$) are used to anchor the Pantheon SNe samples that give the shape of the distance-redshift relation through GP regression. The results are $H_0=72.8_{-1.7}^{+1.6}\rm{\ km/s/Mpc}$ for a flat universe and $H_0=77.3_{-3.0}^{+2.2}\rm{\ km/s/Mpc}$ for a non-flat universe. 
These values are consistent with the results assuming a $\Lambda$CDM model, and have 
comparable uncertainties, though they have the advantage of being cosmology model 
independent (and include SN). With current data, $D_{\rm d}$ 
measurements do not play a significant role, though they 
have the property of being relatable to SN independent of 
spatial curvature. 

We perform several consistency tests of the data, and between the different probes. 
All show consistency, though an odd systematic trend persists in the value of the 
derived $H_0$ with lens redshift. We illustrate this trend through several methods. 
In particular, one could interpret it as a transition at $z\approx0.4$. Irrespective 
of the value of $H_0$, the distances from SL systems lying all below or all above $z\approx0.4$ are highly consistent with the SNe cosmology (this holds for all 7 
such combinations of systems), but comparison of lensing systems on either side of 
$z\approx0.4$ all show an offset (admittedly individually statistically minor) from the 
SN cosmology -- this holds, in the same direction, for all 8 such combinations of systems. 

This could be a statistical fluke (though it has persisted since the first analysis 
with fewer systems) or some observational systematic. We speculate about one possible 
explanation based on physics beyond the standard model, showing how a modification in 
the effect of gravity on light propagation, $\gl(a)$, could account for this. Moreover, 
models in the literature show that the magnitude and redshift dependence of such an effect 
is possible, and this could also affect the perceived value of $\sigma_8$, possibly 
bearing on that tension as well. Given the low statistical significance with current 
data, we merely suggest keeping an eye on whether further, or improved, data continue 
to support such a physics explanation. 

Fortunately time-delay strong lensing is a burgeoning field with the onset of cosmic  
surveys, and more well-measured and well-analyzed lenses are not far off, with imaging 
surveys such as DES, ZTF, LSST, and Euclid. Monitoring campaigns, adaptive optics, and 
spectroscopic followup (including multiobject instruments such as DESI) all play 
important roles as well. More lens systems at all redshifts -- near $z\approx0.4$, 
below, and well above will test whether a standard cosmology matches both strong 
lenses and supernovae. Continued detailed systematics studies of all distance indicators 
-- and all light deflection probes -- will be essential for confirming any result.

\section*{Acknowledgments}

KL was supported by the National Natural Science Foundation of China (NSFC) No.~11973034. AS would like to acknowledge the support of the Korea Institute for Advanced Study (KIAS) grant funded by the Korea government and Kobayashi-Maskawa Institute (KMI) and Nagoya University for their hospitality in the final stages of this project. EL is supported in part by the Energetic Cosmos Laboratory and by the U.S.\ Department of Energy, Office of Science, Office of High Energy Physics, under Award DE-SC-0007867 and contract no. DE-AC02-05CH11231.

\clearpage

\end{document}